\documentclass[epj]{svjour}
\usepackage{graphicx}
\usepackage{amsmath}
\usepackage{bm}
\usepackage{amssymb}
\usepackage{latexsym}
\usepackage{dcolumn}
\usepackage{url}
\newcolumntype{d}{D{.}{.}{-1}}
\begin{document}
\title{Relativistic calculation of L$\beta_2$ satellite spectra of W and Au}
\author{J.\ P.\ Marques\inst{1} \and
        F.\ Parente\inst{1} \thanks{Present address: Departamento de F\'\i sica, Faculdade de Ci\^encias e Tecnologia, Universidade Nova de Lisboa, Monte de Caparica, 2829-516 Caparica, Portugal} \and
        M.\ C.\ Martins\inst{1} \and 
        P.\ Indelicato\inst{2}
%
}                     
\offprints{J.\ P.\ Marques}          
\institute{Centro de F{\'\i}sica At{\'o}mica e Departamento F{\'\i}sica, 
Faculdade de Ci{\^e}ncias, Universidade de Lisboa, \\
Campo Grande, Ed. C8, 1749-016 Lisboa, Portugal, 
\email{jmmarques@fc.ul.pt}
\and 
Laboratoire Kastler Brossel,
\'Ecole Normale Sup\' erieure; CNRS; Universit\' e P. et M. Curie - Paris 6\\
Case 74; 4, place Jussieu, 75252 Paris CEDEX 05, France
\email{paul.indelicato@spectro.jussieu.fr}
}
\date{Received: \today / Revised version: date}
%

\abstract{
L$\beta_2$ X-ray satellite spectra of tungsten and gold are calculated using the Multi-Configuration 
Dirac-Fock energies and compared with recent experimental data.
 New calculations of L$_1$-L$_3$M$_5$ Coster-Kronig transition energies for tungsten are presented,
 confirming the origin of the L$\beta_2$ visible satellites reported by two experimental groups. 
We found the value 5.09 eV for the average energy of the L$_1$-L$_3$M$_5$ Coster-Kronig transition
 in tungsten. A detailed calculation of the L$\beta_2$ and L$\beta_{15}$ satellite spectra was performed for gold.
} 
\maketitle
%

\section{Introduction}
\label{intro}
X-ray lines from almost neutral atoms are seldom pure, being in most cases contaminated by 
the so-called satellite lines, that is, X-ray lines that result 
from transitions in multi-inner hole atomic configurations. In heavy atoms,
X-ray lines resulting from L $\to $ M,N hole transitions with a 
M spectator hole lead, in general, to satellites that can be resolved from 
the diagram line (visible satellites), whereas the ones with N-, O- or P-shell spectator 
holes usually appear embedded within the natural width of the parent line (hidden satellites).
    
The study of X-ray satellite lines in systems where only one spectator 
inner hole exists is more than seventy years old, yet not very many works on 
this subject have been published. Ritchmeyer and 
Ramberg~\cite{1} compared their computed satellite structure 
of L$\alpha_{1,2}$ (L$_{3}$ $\to $ M$_{4,5}$ hole transitions), 
L$\beta_2$ (L$_{3}$ $\to $ N$_{5}$) and
L$\beta_{15}$ (L$_{3}$ $\to $ N$_{4}$) spectra for gold
with experimental data. L$\alpha$ 
X-ray satellite spectra of elements between Zr and Cd were studied by Jusl{\'e}n 
\textit{et al}~\cite{2}. Doyle and Shafroth~\cite{3} measured the
intensities of L$\alpha$ and L$\beta_1$  (L$_{2}$ $\to $ M$_{4}$)
satellites of elements in the 
$37\leq Z\leq 56$ region. Hague \textit{et al}~\cite{4} reported 
on silver L$\alpha$ satellites. Extensive tables of relativistic 
energies of L X-ray satellite lines, using the Dirac-Fock-Slater (DFS) approach 
including quantum electrodynamic corrections, were published by Parente 
\textit{et al}~\cite{5}. The L$\alpha_1$ satellite structures of iridium, gold 
and uranium were studied theoretically by Parente \textit{et al}~\cite{6}  and
experimentally by Carvalho \textit{et al}~\cite{7}. The
possibility that L X-ray satellite lines, not predicted 
by theory, could be present in tungsten was raised by Salgueiro \textit{et al}~\cite{8},
to explain their experimental data. 
A simple model to study the relative intensities of satellite and diagram 
X-ray lines following L-shell ionization in the region around $Z=50$ was
proposed by Xu and Rosato~\cite{9}. Most recently, Vlaicu \textit{et al}~\cite{10}
studied the L emission spectrum of tungsten, confirming 
the existence of the satellite lines found by Salgueiro  
\textit{et al}~\cite{8}, and later Oohashi \textit{et al}~\cite{11} focused on the origin
of the L$\beta_2$ visible satellite lines in gold.

The existence of satellite lines in X-ray spectra has been explained 
long ago by the creation of multiple vacancies in inner-shells, resulting 
from the ionization process. Two mechanisms have been proposed for the 
creation of double vacancies after a primary ionization in the L$_1$ or L$_2$ subshells, 
namely shake-off and L$_i$-L$_jX_k$ Coster-Kronig transitions.
Here $i=1,2$, $j=2,3$ (with $i\neq j$), and $X_k$ may be any outer subshell. The shake 
process, for energies not far above the threshold, depends on the
excitation energy~\cite{12}, whereas Coster-Kronig transitions are 
independent on the excitation energy. Furthermore, Coster-Kronig 
transitions are highly probable when energetically allowed so that, 
in this case, the atom, after the primary ionization, will 
unavoidably end up with two inner-shell vacancies.

Atomic K- and L-shells relativistic radiationless transition probabilities for  
22 elements with atomic numbers $18\leq Z \leq 96$ were published 
in 1979 by Chen \textit{et al}~\cite{13}. Earlier, in 1977, the same authors had
found L$_1$-L$_3$M$_{4,5}$ Cos\-ter-Kronig transitions to be energetically 
forbidden in the region $50 \leq Z \leq 79$~\cite{14}. However, 
in 1987, Salgueiro \textit{et al}~\cite{8} obtained tungsten L$\alpha$ X-ray 
spectra where satellite lines were present, which could not be explained 
by the shake mechanism. Therefore they suggested that, most probably,
Coster-Kronig L$_1$-L$_3$M$_5$ transitions could take place in tungsten, 
contrary to Chen \textit{et al} predictions. The existence of these 
satellite lines was confirmed in 1998 by Vlaicu and co-workers~\cite{10}. 
We note, however, that Agarwal~\cite{15} already referred the existence 
of L$_1$-L$_3$M$_5$ Coster-Kronig transitions for $Z > 73$ in 1979.
      
In this work we used the multi-configuration Dirac-Fock code of 
Desclaux and Indelicato~\cite{16,17,17a} to calculate the energies of L$_1$- 
and L$_3$M$_{4,5}$-hole levels of tungsten, looking for the possibility of existence 
of L$_1$-L$_3$M$_{4,5}$ Coster-Kronig transitions. Our results led to the 
conclusion that L$_1$-L$_3$M$_5$ transitions are indeed energetically allowed in tungsten, 
contrary to Chen \textit{et al}~\cite{14} prediction. Furthermore, 
we used the same code to compute transition energies of the X-ray lines 
in the L$\beta_2$ satellite band of tungsten. As the number of lines involved 
is too big to handle, we used the method suggested by 
Parente \textit{et al}~\cite{5}, using Racah's algebra, to predict the 
shape of this satellite band.

The L$\beta_2$ and L$\beta_{15}$ satellite bands of gold, where the number 
of transitions allows for a full calculation, were also 
studied in this work and compared with recent experimental 
results~\cite{11}. Spectator holes in all possible 
subshells were considered in the study, including the ones that give origin to 
hidden satellites. To assess the accuracy of the method that makes use of 
Racah's algebra, we also used this method for gold, to allow for a comparison
with the results of the full calculation.

\section{Calculation of atomic wave functions and transition probabilities}
\label{sec:1}

Bound states wave functions are calculated using the Di\-rac-Fock program of J.
P. Desclaux and P. Indelicato. Details on the Hamiltonian and the processes
used to build the wave-functions can be found elsewhere~\cite{16,17,18}.

The total wave function is calculated with the help of the variational
principle. The total energy of the atomic system is the eigenvalue of the
equation
\begin{equation}
\label{eq001}
        {\cal H}^{\mbox{{\tiny no pair}}}
        \Psi_{\Pi,J,M}(\ldots,\bm{r}_{i},\ldots)=E_{\Pi,J,M}
        \Psi_{\Pi,J,M}(\ldots,\bm{r}_{i},\ldots),
\end{equation}
where $\Pi$ is the parity, $J$ is the total angular momentum eigenvalue, and
$M$ is the eigenvalue of its projection on the $z$ axis $J_{z}$. The MCDF
method is defined by the particular choice of a trial function to solve equation
(\ref{eq001}) as a linear combination of configuration state functions (CSF):
\begin{equation}
\left\vert \mathit{\Psi}_{\mathit{\Pi},J,M}\right\rangle =\sum_{\nu=1}%
^{n}c_{\nu}\left\vert \nu,\mathit{\Pi},J,M\right\rangle . \label{eq_cu}%
\end{equation}
The CSF are also eigenfunctions of the parity $\mathit{\Pi}$, the total
angular momentum $J^{2}$ and its projection $J_{z}$. The label $\nu$ stands
for all other numbers (principal quantum number, ...) necessary to define
unambiguously the CSF. The $c_{\nu}$ are called the mixing coefficients and
are obtained by diagonalization of the Hamiltonian matrix coming from the
minimization of the energy in equation~(\ref{eq_cu}) with respect to the $c_{\nu}%
$.
The CSF are antisymmetric products of one-electron wave functions
expressed as linear combination of Slater determinants of Dirac
4-spinors

\begin{equation}
\left\vert \nu,\mathit{\Pi},J,M\right\rangle =\sum_{i=1}^{N_{\nu}}%
d_{i}\left\vert
\begin{array}
[c]{ccc}%
\psi_{1}^{i}\left(  r_{1}\right)  & \cdots & \psi_{m}^{i}\left(  r_{1}\right)
\\
\vdots & \ddots & \vdots\\
\psi_{1}^{i}\left(  r_{m}\right)  & \cdots & \psi_{m}^{i}\left(  r_{m}\right)
\end{array}
\right\vert .
\end{equation}
where the $\psi$-s are the one-electron wave functions and the coefficients
$d_{i}$ are determined by requiring that the CSF is an eigenstate of $J^{2}$
and $J_{z}$.
The $d_{i}$ coefficients are obtained by requiring that the CSF are
eigenstates of $J^{2}$ and $J_{z}.$
A variational principle provides the integro-differential equations to
determine the radial wave functions and a Hamiltonian matrix that provides the
mixing coefficients $c_{\nu}$ by diagonalization. One-electron radiative
corrections (self-energy and vacuum polarization) are added afterwards. All
the energies are calculated using the experimental nuclear charge distribution
for the nucleus.

The so-called Optimized Levels (OL) method was used to determine the wave
function and energy for each state involved. Thus, spin-orbitals in the
initial and final states are not orthogonal, since they have been optimized
separately. The formalism to take in account the wave functions
non-orthogonality in the transition probabilities calculation has been
described by L\"{o}wdin~\cite{19}. The matrix element of a one-electron
operator $O$ between two determinants belonging to the initial and final
states can be written
\begin{align}
&  \left\langle \nu\mathit{\Pi}JM\right\vert \sum_{i=1}^{N}O\left(
r_{i}\right)  \left\vert \nu^{\prime}\mathit{\Pi}^{\prime}J^{\prime}M^{\prime
}\right\rangle = \nonumber\\
& \times  \frac{1}{N!}   \left\vert
\begin{array}
[c]{ccc}%
\psi_{1}\left(  r_{1}\right)  & \cdots & \psi_{m}\left(  r_{1}\right) \\
\vdots & \ddots & \vdots\\
\psi_{1}\left(  r_{m}\right)  & \cdots & \psi_{m}\left(  r_{m}\right)
\end{array}
\right\vert \nonumber \\
& \times \sum_{i=1}^{m}O\left(  r_{i}\right)  \left\vert
\begin{array}
[c]{ccc}%
\phi_{1}\left(  r_{1}\right)  & \cdots & \phi_{m}\left(  r_{1}\right) \\
\vdots & \ddots & \vdots\\
\phi_{1}\left(  r_{m}\right)  & \cdots & \phi_{m}\left(  r_{m}\right)
\end{array}
\right\vert , \label{eq002}
\end{align}where the $\psi_{i}$ belong to the initial state and the $\phi_{i}$ and primes
belong to the final state. If $\psi=\left\vert n\kappa\mu\right\rangle $ and
$\phi=\left\vert n^{\prime}\kappa^{\prime}\mu^{\prime}\right\rangle $ are
orthogonal, i.e., $\left\langle n\kappa\mu|n^{\prime}\kappa^{\prime}%
\mu^{\prime}\right\rangle =\delta_{n,n^{\prime}}\delta_{\kappa,\kappa^{\prime
}}\delta_{\mu,\mu^{\prime}}$, the matrix element (\ref{eq002}) reduces to one
term $\left\langle \psi_{i}\right\vert O\left\vert \phi_{i}\right\rangle $
where $i$ represents the only electron that does not have the same
spin-orbital in the initial and final determinants. Since $O$ is a
one-electron operator, only one spin-orbital can change, otherwise the matrix
element is zero. In contrast, when the orthogonality between initial and final
states is not enforced, one gets~\cite{19}
\begin{equation}
\left\langle \nu\mathit{\Pi}JM\right\vert \sum_{i=1}^{N}O\left(  r_{i}\right)
\left\vert \nu^{\prime}\mathit{\Pi}^{\prime}J^{\prime}M^{\prime}\right\rangle
=\sum_{i,j^{\prime}}\left\langle \psi_{i}\right\vert O\left\vert
\phi_{j^{\prime}}\right\rangle D_{ij^{\prime}},
\end{equation}
where $D_{ij^{\prime}}$ is the minor determinant obtained by crossing out the
$i$th row and $j^{\prime}$th column from the determinant of dimension $N\times
N$, made of all possible overlaps $\left\langle \psi_{k}|\phi_{l^{\prime}%
}\right\rangle $.

Radiative corrections are also introduced, from a full QED treatment. The
one-electron self-energy is evaluated using the one-electron values of Mohr
and coworkers~\cite{20,21,22} and corrected for finite nuclear size~\cite{23}. 
The self-energy screening and vacuum polarization are treated with an approximate 
method developed by Indelicato and co\-workers~\cite{24,25,26,27}.

\section{Results}
\label{sec:2}

\subsection{Tungsten}
\label{sec:2.1}

Using the MCDF code of Desclaux and Indelicato~\cite{16,17}, we calculated
the energies of all levels in
the L$_{1}$- and L$_{3}$M$_{5}$-hole configurations of
tungsten, to check for the possibility of Coster-Kronig transitions between
these two configurations. Interaction of the inner holes with electrons in outer
unfilled shells was taken in account and full relaxation was included but no electronic correlation.

There are 63 energy levels in the L$_{1}$ configuration. We
found 359 energy levels in the L$_{3}$M$_{5}$ configuration with
energies below the lowest energy level of the L$_{1}$ configuration. Thus, we
arrive to the conclusion that L$_{1}$-L$_{3}$M$_{5}$ Coster-Kronig transitions are indeed
possible in tungsten, originating the double L$_{3}$M$_{5}$ vacancy state which afterwards 
yield the satellite lines first detected by
Salgueiro \textit{et al}~\cite{8} and confirmed
by Vlaicu \textit{et al}~\cite{10}. Averaging over initial and final
state energies, we found for the radiationless transition the energy of 5.09
eV.

This value is to be compared to -2.70 eV found by Chen \textit{et al}~\cite{14}, 
using a relativistic DFS approach. In the latter work coupling with
unfilled outer shells was neglected. These authors present the Coster-Kronig
transition energies as differences of initial and final average total system energies,
which does not allow for a detailed comparison between individual level energies.

Using the MCDF computer code, we calculated the energies of all possible
L$_{3}\to $N$_5$ radiative transitions in the presence of a M$_{5}$ spectator hole,
L$_{3}$M$_{5}$-M$_{5}$N$_5$ transitions, that yield the L$\beta_{2}$
satellite band. In general, to each pair of initial and final
angular momenta correspond several X-ray transitions. A full calculation of transition 
probabilities would be, in this case, a formidable task, due to the enormous
number, more than one hundred thousand, of authorized transitions. Instead, we used
Racah's algebra to compute the relative line intensities, using the
method of Parente \textit{et al}~\cite{5}. This method assumes:

\begin{enumerate}
\item The initial multiplet states are populated statistically;

\item The effect of the multiplet energy differences on the X-ray matrix
elements can be neglected;

\item The Auger decay rates of the various multiplet states of a configuration
can be taken to be identical. In fact, for heavy atoms there are so many open
Auger channels that the multiplet effect on Auger emission rates for
double-hole states becomes minimal.
\end{enumerate}

The angular momenta resulting from the unfilled shells are $j_1, j_2, j_3$ 
and $j'_1, j'_2, j'_3$ for the initial and final configurations, respectively, 
where ${j_{3}}$ and ${j_{3}^{\prime}}$ refer to the outer shell electrons.

Under the above assumptions, the intensity for the line
\begin{center}
$\left\vert ~j_{1}j_{2}(J_{12}),j_{3};J\right\rangle \to\left\vert
j_{1}^{\prime}j_{2}^{\prime}(J_{12}^{\prime}),~j_{3}^{\prime};J^{\prime
}\right\rangle $
\end{center}
is found to be proportional to
\begin{align}
&  \left\vert \underset{J_{13}^{\prime}}{\sum}\left(  2J_{13}^{\prime
}+1\right) \sqrt{\left(  2J_{12}^{\prime}+1\right)  \left(  2J_{12}+1\right)
\left(  2J^{\prime}+1\right)  \left(  2J+1\right)} \right. \nonumber\\
& \left. \times \sqrt{\left(  2j_{1}+1\right)\left(  2j_{2}^{\prime}+1\right)}  \right. \nonumber \\
& \times \left. \left(  -1\right)  ^{J_{12}^{\prime}+j_{3}^{\prime} +3J_{13}^{\prime}+j_{2}^{\prime}+3j_{1}+j_{2}+j_{3}+J^{\prime}+1} \right. \nonumber\\
& \left. \times \left\{\begin{array}
[c]{ccc}%
J_{12}^{\prime} & j_{3}^{\prime} & J^{\prime}\\
J_{13}^{\prime} & j_{2}^{\prime} & j_{1}^{\prime}%
\end{array}
\right\}  \left\{
\begin{array}
[c]{ccc}%
j_{1} & j_{2} & J_{12}\\
j_{3} & J & J_{13}^{\prime}%
\end{array}
\right\}  \left\{
\begin{array}
[c]{ccc}%
j_{2}^{\prime} & J^{\prime} & J_{13}^{\prime}\\
J & j_{1} & 1
\end{array}
\right\}  \right\vert ^{2}, \label{eq003}%
\end{align}
where $J_{ik}$ results from the coupling between angular momenta $j_{i}$ and $j_{k}$.
In what concerns the L$_{3}$M$_{5}$-M$_{5}$N$_{5}$ transitions, we have
$j_{1}=\dfrac{3}{2}$; $j_{2}=j_{2}^{\prime}=\dfrac{5}{2}$; $j_{1}^{\prime
}=\dfrac{5}{2}$. For tungsten $j_{3}=0,\ldots,6$; $j_{3}^{\prime}=0,\ldots,6$;
$J=0,\ldots,10$; $J^{\prime}=0,\ldots,11$. For each pair of $J,J^{\prime}$ values,
 the remaining angular momenta values can be found using
standard angular momenta calculations. We assumed only one line for each
pair of initial and final angular momenta values, assigning to this line the weighted
energy and the sum  of the line component intensities, in arbitrary units.
As L$_{3}$M$_{5}$-M$_{5}$N$_{4}$ transition lines are much less intense 
that L$_{3}$M$_{5}$-M$_{5}$N$_{5}$ lines, we did not include them
in the calculation for tungsten.

\begin{figure}[h]
\centering
\includegraphics[width=8.8cm]{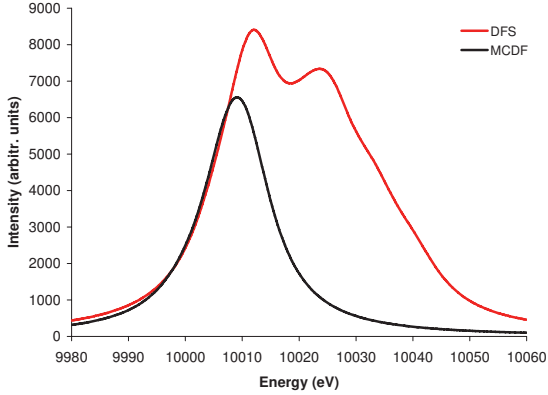}
\caption{Calculated L$\beta_2$ satellite band of tungsten,
using Dirac-Fock-Slater - DFS~\cite{5},
and Multi-Configuration
Dirac-Fock - MCDF (this work) energy values, respectively.
In both cases, relative intensities were calculated in the present work
using Racah's algebra, as described in the text.
The bands are normalized in intensity.}
\label{fig01}
\end{figure}

We assume that the natural width of the L$_{3}$M$_{5}$-M$_{5}$N$_{5}$ satellite
line is given by ~\cite{14}
\begin{equation}
\Gamma\left(  \text{L}_{3}\text{M}_{5}\text{-M}_{5}%
\text{N}_{5}\right)  =\Gamma _{\text{L}_{3}}^{\text{tot}}+2\Gamma _{\text{M}_{5}}^{\text{tot}} +\Gamma _{\text{N}_{5}}^{\text{tot}},
\end{equation}
where $\Gamma _{X}^{\text{tot}}$ is the natural width of $X$ level.
Using the  values proposed by Campbell and Papp \cite{30} we obtained 12.2 eV 
for the width of  each of those satellite lines. In this way we arrived at the
tungsten L$\beta_2$ satellite band presented in figure~\ref{fig01}, together with the corresponding
spectrum generated using the DFS energy values of Parente \textit{et al}~\cite{5}.

Due to the fact that only the lowest levels of the initial L$_3$M$_5$ hole configuration are
fed by Coster-Kronig transitions from the L$_1$ hole initial configuration,
the theoretical X-ray satellite band  generated in this work is much narrower than 
if all initial levels were fed. This is clearly shown in figure~\ref{fig01}, as
the satellite band computed using DFS energy values included all energy levels 
of the L$_3$M$_5$ configuration as initial levels.

This prediction can be
compared with the experimental data of Vlaicu \textit{et al} (figure 2 of ~\cite{10}).
Although the data resolution is very poor, for obvious
lack of statistics, the proeminent peak with energy around 10010 eV is 
well reproduced in the theoretical band (figure~\ref{fig01} - MCDF). It is obvious that experimental results
with better statistics are needed.

Vlaicu \textit{et al}~\cite{10} estimated relative intensities of the L-satellite lines for tungsten,
using the expressions proposed by Xu and Rosato~\cite{9}, following the
model of Krause \textit{et al }~\cite{31},
with values obtained by other authors for the pertinent quantities.
As they used Chen \textit{et al}~\cite{14} results for
the Coster-Kronig transition probabilities, they assumed that the
Coster-Kronig channel for creating M-shell spectator holes is closed, contrary
to the findings of the present work.


\subsection{Gold}
\label{sec:3}

\subsubsection{Coster-Kronig transition energies}
\label{sec:3.1}

The L$\beta_{2}$ and L$\beta_{15}$ satellite bands of gold were  studied
in this work. In this case, the existence of both L$_{1}$-L$_{3}$M$_{4}$ and
L$_{1}$-L$_{3}$M$_{5}$ Coster-Kronig transitions were predicted by Chen
\textit{et al }~\cite{14} and confirmed by our own calculation.

We found for the L$_{1}$-L$_{3}$M$_{4}$ Coster-Kronig transition in gold the
energy of $38.23$ eV and for the L$_{1}$-L$_{3}$M$_{5}$ transition the energy
of $132.15$ eV. These values are close to the  37.65 eV and 130.39 eV values,
respectively, obtained by Chen \textit{et al}~\cite{13}.

The theoretical L$\beta_{2}$
satellite band of gold was obtained by calculating the transition
probabilities for all possible dipolar electric transitions corresponding to
the decay of the L$_{3}$M$_{5}$ and L$_{3}$M$_{4}$ double-hole configurations,
in the energy window between 11600-11700 eV, and, in a separate way, using
Racah's algebra method, as was done for tungsten.

\subsubsection{Full MCDF calculation of satellite band}
\label{sec:3.2}

In gold, 8 and 7 energy levels correspond to the L$_{3}$M$_{5}$- and L$_{3}$M$_{4}$-hole configurations
 respectively. On the other hand,  11 and 8 levels correspond to the M$_{5}$%
N$_{5}$ and M$_{4}$N$_{5}$ configurations. This leads to 149 possible electric-dipole allowed lines
between the levels of the L$_{3}$M$_{4,5}$ and M$_{4,5}$N$_{5}$ configurations
that yield the L$\beta_{2}$ visible satellite band.
The calculations of the energies and transition probabilities for these lines
were performed using the MCDF code of Desclaux and Indelicato, in
single-configuration mode, which includes all relativistic CSF originated from a
single LS configuration. 

\begin{table*}[t]
\begin{center}
\caption{Energies in eV and transition probabilities ($W$)
in s$^{-1}$ for gold L$\beta_2$, L$\beta_{15}$, and L$\beta_3$ diagram
lines.}
\label{diagram}
\begin{tabular}{cc|cc|cc}\hline \hline
\multicolumn{2}{c|}{L$_3$ - N$_5$ (L$\beta_2$)} & 
\multicolumn{2}{c|}{L$_3$ - N$_4$ (L$\beta_{15}$)} &
\multicolumn{2}{c} {L$_1$ - M$_3$ (L$\beta_3$)} \\ \hline 
Energy    &      $W$             &    Energy   &      $W$                &  Energy   &      $W$   \\
11581.36 & 1.38$\times 10^{14}$ & 11563.24 & 1.28$\times 10^{13}$ & 11612.24 & 1.67$\times 10^{14}$ \\
    11581.33 & 1.54$\times 10^{13}$ & 11563.20 & 2.66$\times 10^{12}$ & 11611.82 & 8.31$\times 10^{13}$ \\
    11581.40 & 2.14$\times 10^{14}$ & 11563.21 & 2.55$\times 10^{12}$ & 11611.87 & 4.15$\times 10^{14}$ \\
              &                      &  11563.17 &  2.28$\times 10^{13}$ \\
\hline
\end{tabular}
\end{center} 
\end{table*}

\begin{table*}
\begin{center}
\caption{Energies in eV and transition probabilities ($W$)
in s$^{-1}$ for gold L$\beta_2$ and L$\beta_{15}$ visible satellite lines.}
\label{satellite}
\begin{tabular}{cc|cc|cc|cc}\hline \hline
\multicolumn{2}{c|}{L$_3$M$_5$ - M$_5$N$_5$} & 
\multicolumn{2}{c|}{L$_3$M$_4$ - M$_4$N$_5$} &
\multicolumn{2}{c|}{L$_3$M$_5$ - M$_5$N$_4$} & 
\multicolumn{2}{c} {L$_3$M$_4$ - M$_4$N$_4$} \\ \hline 
Energy    & $W$         & Energy    & $W$ &  Energy   & $W$ & Energy   & $W$ \\
11627.05 & 3.38$\times 10^{06}$ & 11645.44 & 5.88$\times 10^{11}$ & 11614.35 &  4.44$\times 10^{12}$ & 11616.27 &  4.51$\times 10^{07}$ \\
11630.44 & 3.69$\times 10^{11}$ & 11645.50 & 1.12$\times 10^{12}$ & 11614.37 &  9.63$\times 10^{10}$ & 11625.33 &  4.12$\times 10^{12}$ \\
11630.50 & 1.25$\times 10^{13}$ & 11645.52 & 9.47$\times 10^{10}$ & 11614.41 &  1.13$\times 10^{11}$ & 11625.34 &  2.94$\times 10^{11}$ \\
11631.65 & 1.28$\times 10^{13}$ & 11646.48 & 3.27$\times 10^{08}$ & 11614.43 &  3.57$\times 10^{12}$ & 11625.35 &  3.04$\times 10^{11}$ \\
11631.71 & 2.85$\times 10^{11}$ & 11647.89 & 9.95$\times 10^{11}$ & 11615.88 &  6.82$\times 10^{08}$ & 11625.35 &  2.70$\times 10^{12}$ \\
11635.82 & 1.91$\times 10^{13}$ & 11647.91 & 9.18$\times 10^{12}$ & 11617.83 &  2.24$\times 10^{12}$ & 11625.97 &  1.95$\times 10^{12}$ \\
11635.84 & 1.34$\times 10^{12}$ & 11647.94 & 1.45$\times 10^{13}$ & 11617.89 &  1.73$\times 10^{12}$ & 11625.98 &  9.88$\times 10^{11}$ \\
11635.90 & 1.37$\times 10^{12}$ & 11647.96 & 8.51$\times 10^{11}$ & 11617.89 &  6.45$\times 10^{10}$ & 11629.82 &  1.39$\times 10^{08}$ \\
11635.93 & 1.22$\times 10^{13}$ & 11650.18 & 2.27$\times 10^{12}$ & 11621.86 &  1.38$\times 10^{12}$ & 11633.24 &  2.47$\times 10^{12}$ \\
11638.60 & 1.16$\times 10^{12}$ & 11650.20 & 3.49$\times 10^{13}$ & 11621.90 &  2.44$\times 10^{12}$ & 11633.25 &  1.27$\times 10^{11}$ \\
11638.66 & 5.08$\times 10^{13}$ & 11650.24 & 4.96$\times 10^{13}$ & 11621.92 &  2.86$\times 10^{11}$ & 11633.26 &  1.75$\times 10^{12}$ \\
11638.70 & 6.23$\times 10^{13}$ & 11653.43 & 1.53$\times 10^{08}$ & 11623.20 &  1.65$\times 10^{08}$ & 11634.52 &  2.70$\times 10^{12}$ \\
11639.27 & 5.04$\times 10^{05}$ & 11654.39 & 2.35$\times 10^{12}$ & 11624.62 &  2.74$\times 10^{10}$ & 11634.54 &  3.13$\times 10^{11}$ \\
11640.46 & 1.06$\times 10^{12}$ & 11654.41 & 6.01$\times 10^{13}$ & 11624.65 &  1.72$\times 10^{11}$ & 11634.55 &  1.51$\times 10^{12}$ \\
11640.46 & 1.38$\times 10^{12}$ & 11654.45 & 7.81$\times 10^{13}$ & 11624.65 &  2.57$\times 10^{11}$ & 11634.63 &  2.90$\times 10^{12}$ \\
11640.52 & 4.36$\times 10^{10}$ & 11655.15 & 2.25$\times 10^{12}$ & 11624.68 &  1.35$\times 10^{10}$ & 11634.69 &  1.43$\times 10^{11}$ \\
11646.62 & 3.07$\times 10^{12}$ & 11655.15 & 4.37$\times 10^{12}$ & 11626.66 &  1.25$\times 10^{11}$ & 11634.71 &  2.00$\times 10^{12}$ \\
11646.69 & 5.16$\times 10^{12}$ & 11655.22 & 1.14$\times 10^{13}$ & 11626.66 &  2.75$\times 10^{12}$ & 11635.05 &  1.83$\times 10^{12}$ \\
11646.69 & 2.86$\times 10^{12}$ & 11655.23 & 2.12$\times 10^{12}$ & 11626.68 &  1.91$\times 10^{12}$ & 11635.06 &  1.95$\times 10^{11}$ \\
11646.71 & 5.90$\times 10^{11}$ & 11655.82 & 4.42$\times 10^{11}$ & 11633.46 &  5.46$\times 10^{08}$ & 11635.06 &  1.00$\times 10^{12}$ \\
11647.32 & 3.55$\times 10^{12}$ & 11655.85 & 6.54$\times 10^{11}$ & 11634.69 &  5.01$\times 10^{12}$ & 11638.91 &  6.71$\times 10^{06}$ \\
11647.38 & 1.72$\times 10^{11}$ & 11655.87 & 1.36$\times 10^{10}$ & 11634.70 &  1.46$\times 10^{11}$ & 11642.44 &  1.14$\times 10^{07}$ \\
11647.46 & 2.46$\times 10^{12}$ & 11657.61 & 3.27$\times 10^{12}$ & 11634.76 &  3.81$\times 10^{12}$ & 11642.54 &  6.92$\times 10^{12}$ \\
11649.22 & 1.05$\times 10^{12}$ & 11657.62 & 1.74$\times 10^{13}$ & 11636.15 &  3.11$\times 10^{12}$ & 11642.55 &  3.05$\times 10^{11}$ \\
11649.25 & 1.49$\times 10^{13}$ & 11657.67 & 3.08$\times 10^{13}$ & 11636.18 &  2.22$\times 10^{12}$ & 11642.60 &  2.61$\times 10^{11}$ \\
11649.29 & 2.13$\times 10^{13}$ & 11658.09 & 4.67$\times 10^{11}$ & 11636.21 &  1.80$\times 10^{11}$ & 11642.62 &  5.18$\times 10^{12}$ \\
11650.77 & 1.21$\times 10^{12}$ & 11658.11 & 9.65$\times 10^{12}$ & 11638.16 &  1.01$\times 10^{11}$ & 11644.24 &  9.65$\times 10^{11}$ \\
11650.83 & 3.16$\times 10^{13}$ & 11658.15 & 1.30$\times 10^{13}$ & 11638.17 &  2.10$\times 10^{12}$ & 11644.25 &  2.05$\times 10^{11}$ \\
11651.99 & 3.29$\times 10^{13}$ & 11658.17 & 5.34$\times 10^{11}$ & 11638.22 &  1.57$\times 10^{12}$ & 11644.26 &  2.09$\times 10^{11}$ \\
11658.25 & 1.85$\times 10^{07}$ & 11659.00 & 8.31$\times 10^{12}$ & 11638.22 &  9.85$\times 10^{10}$ & 11644.26 &  4.15$\times 10^{11}$ \\
11658.93 & 7.78$\times 10^{07}$ & 11659.07 & 1.63$\times 10^{13}$ & 11646.35 &  9.90$\times 10^{10}$ & 11644.42 &  5.92$\times 10^{06}$ \\
11660.33 & 1.06$\times 10^{13}$ & 11659.91 & 6.87$\times 10^{09}$ & 11646.36 &  2.29$\times 10^{11}$ & 11648.09 &  1.00$\times 10^{12}$ \\
11660.42 & 1.19$\times 10^{12}$ & 11661.46 & 1.57$\times 10^{09}$ & 11646.41 &  5.56$\times 10^{11}$ & 11648.11 &  4.67$\times 10^{11}$ \\
11660.43 & 5.88$\times 10^{12}$ & & & 11646.42 &  1.18$\times 10^{11}$ & & \\
11660.73 & 6.33$\times 10^{11}$ & & & 11649.14 &  1.94$\times 10^{11}$ & & \\
11660.78 & 1.27$\times 10^{13}$ & & & 11649.15 &  9.40$\times 10^{11}$ & & \\
11660.80 & 1.71$\times 10^{13}$ & & & 11649.17 &  1.75$\times 10^{12}$ & & \\
11660.85 & 6.47$\times 10^{11}$ & & & 11651.17 &  3.52$\times 10^{08}$ & & \\
11671.11 & 2.33$\times 10^{12}$ & & & & & & \\
11671.12 & 4.69$\times 10^{12}$ & & & & & & \\
11671.18 & 2.39$\times 10^{12}$ & & & & & & \\
11671.19 & 4.81$\times 10^{12}$ & & & & & & \\
11671.20 & 1.20$\times 10^{13}$ & & & & & & \\
11671.21 & 2.41$\times 10^{12}$ & & & & & & \\
11673.74 & 4.37$\times 10^{07}$ & & & & & & \\
\hline
\end{tabular}
\end{center} 
\end{table*}

The L$\beta_{15}$ (L$_{3}$M$_{4,5}%
$-M$_{4,5}$N$_{4}$) visible satellite lines, although much less
intense than the previous ones, also exist in the same energy window. So, we
included these lines in our calculation.

Results for transition energies and probabilities
of L$\beta_{2}$, L$\beta_{15}$ and
L$\beta_{3}$ (L$_{1}$ $\to $ M$_{3}$) diagram lines
are presented on Table \ref{diagram}. Transition energies and probabilities for
L$\beta_{2}$ and  L$\beta_{15}$ visible satellite lines are shown on Table \ref{satellite}.

Furthermore, satellite lines originating 
from double-holes, in L$_{3}$ and N, O, or P subshells, have energies that are embedded
in the natural widths of the diagram lines. So, the corresponding transition energies and probabilities have also 
been calculated.

In order to compare MCDF calculations of transition energies and probabilities
with experiment, we have to know the X-ray production cross
sections for lines arising from single- and double-hole states in gold. Again, the
pertinent expressions can be found in Xu and Rosato~\cite{9}, using the
model of Krause \textit{et al}~\cite{31}.

We define the cross
section for creation of a single-hole configuration in the L$_{i}$ subshell 
as
%
\begin{equation}
\sigma_{\text{L}_{i}}^{\prime}  =\sigma_{\text{L}_{i}}\left(  1-Q_{\text{L}_{i}}\right),\label{eq051}
\end{equation}
whereas for the L$_{3}X$ $\left(  X=\text{M,N,O,P}\right)  $ double-hole
configurations the corresponding expressions are%
\begin{eqnarray}
\sigma _{\text{L}_{3}X}^{\prime } &=&\sigma _{\text{L}_{1}}\left( 1-Q_{\text{L}_{1}}\right)
f_{13}P\left( \text{L}_{1}\text{-L}_{3}X\right)   \notag \\
&&+\ \sigma _{\text{L}_{2}}\left( 1-Q_{\text{L}_{2}}\right) f_{23}P\left( \text{L}_{2}\text{-L}%
_{3}X\right) \notag \\
&&+\ \sigma _{3}Q_{\text{L}_{3}}\left(
X\right)  .  \label{eq052}
\end{eqnarray}
In equations (\ref{eq051}) and (\ref{eq052}) $\sigma_{\text{L}_{i}}$ are the L$_{i}$-subshell
ionization cross sections, $Q_{\text{L}_{i}}$ is the sum of the shake-off probabilities from all
possible orbitals when a hole is created in the L$_{i}$ subshell, $f_{ij}$ is the
partial Coster-Kronig transition probability from level
L$_{i}$ to level L$_{j}$, $P\left(  \text{L}_{i}\text{-L}_{j}X\right)  $ is the
relative probability of the radiationless transition L$_{i}$-L$_{j}X$
that results in the double vacancy state L$_{j}X$, and $Q_{\text{L}_{3}}\left(
X\right)  $ is the probability of shake-off from the $X$ orbital when a hole
is created in the L$_3$ subshell. It is
assumed that L$_{i}$ vacancies always decay before the outer M holes.

Ionization cross sections are taken from P\'{a}linkas and Schlenk~\cite{32},
for 60 keV electron impact, and shake-off probabilities are
calculated in this work with MCDF wavefunctions (Table ~\ref{sigmas}). The quantities
$P\left(  \text{L}_{i}\text{-L}_{3}X\right)  $ in equation (\ref{eq052}) are taken
from Chen \textit{et al}~\cite{13} corresponding values for Hg, corrected for the difference in atomic number, and Coster-Kronig transition probabilities are from
Chen \textit{et al}~\cite{14}. Values of $\sigma _{\text{L}_{3}X}^{\prime }$
calculated in this way are presented in the last column of Table~\ref{sigmas}.

\begin{table*}
\begin{center}
\caption{Shake-off probabilities from the $X$ orbital when a hole is created in the L$_i$ subshell (this work) and cross sections for creation of double L$_{3}X$ hole configurations in barn, for gold.}
\label{sigmas}
\begin{tabular}{crrrr}\hline \hline
$X$ & $Q_{L_1}(X)$&$Q_{L_2}(X)$&$Q_{L_3}(X)$&$\sigma'_{L_3X}$ \\ 
\hline
K & 2.82$\times 10^{-8}$ & 1.65$\times 10^{-7}$ & 3.73$\times 10^{-8}$ & \\
L$_1$ & 1.11$\times 10^{-5}$ & 1.51$\times 10^{-5}$ & 1.02$\times 10^{-5}$ & \\
L$_2$ & 1.25$\times 10^{-5}$ & 2.69$\times 10^{-5}$ & 1.38$\times 10^{-5}$ & \\
L$_3$ & 3.05$\times 10^{-5}$ & 5.49$\times 10^{-5}$ & 3.77$\times 10^{-5}$ & \\
M$_1$ & 8.95$\times 10^{-5}$ & 1.36$\times 10^{-4}$ & 1.20$\times 10^{-4}$ & \\
M$_2$ & 1.63$\times 10^{-4}$ & 1.53$\times 10^{-4}$ & 1.80$\times 10^{-4}$ & \\
M$_3$ & 3.60$\times 10^{-4}$ & 4.56$\times 10^{-4}$ & 3.51$\times 10^{-4}$ & \\
M$_4$ & 5.26$\times 10^{-4}$ & 4.95$\times 10^{-4}$ & 5.89$\times 10^{-4}$ & 24.1 \\
M$_5$ & 7.74$\times 10^{-4}$ & 8.93$\times 10^{-4}$ & 7.52$\times 10^{-4}$ & 31.9 \\
N$_1$ & 3.99$\times 10^{-4}$ & 4.92$\times 10^{-4}$ & 4.68$\times 10^{-4}$ & 3.0 \\
N$_2$ & 6.71$\times 10^{-4}$ & 6.37$\times 10^{-4}$ & 7.12$\times 10^{-4}$ & 7.6 \\
N$_3$ & 1.44$\times 10^{-3}$ & 1.63$\times 10^{-3}$ & 1.42$\times 10^{-3}$ & 3.8 \\
N$_4$ & 2.81$\times 10^{-3}$ & 2.62$\times 10^{-3}$ & 2.98$\times 10^{-3}$ & 11.4 \\
N$_5$ & 4.23$\times 10^{-3}$ & 4.41$\times 10^{-3}$ & 4.16$\times 10^{-3}$ & 8.2 \\
N$_6$ & 9.19$\times 10^{-3}$ & 8.87$\times 10^{-3}$ & 9.13$\times 10^{-3}$ & 7.6 \\
N$_7$ & 1.22$\times 10^{-2}$ & 1.19$\times 10^{-2}$ & 1.21$\times 10^{-2}$ & 9.9 \\
O$_1$ & 1.89$\times 10^{-3}$ & 2.02$\times 10^{-3}$ & 2.01$\times 10^{-3}$ & 1.7 \\
O$_2$ & 3.38$\times 10^{-3}$ & 3.34$\times 10^{-3}$ & 3.46$\times 10^{-3}$ & 3.2 \\
O$_3$ & 7.92$\times 10^{-3}$ & 9.07$\times 10^{-3}$ & 7.83$\times 10^{-3}$ & 4.9 \\
O$_4$ & 4.11$\times 10^{-2}$ & 4.05$\times 10^{-2}$ & 4.19$\times 10^{-2}$ & 24.2 \\
O$_5$ & 7.44$\times 10^{-2}$ & 7.61$\times 10^{-2}$ & 7.38$\times 10^{-2}$ & 41.4 \\
P$_1$ & 7.72$\times 10^{-2}$ & 7.67$\times 10^{-2}$ & 7.66$\times 10^{-2}$ & 41.4 \\
\hline
Total ($Q$) & 2.39$\times 10^{-1}$ & 2.41$\times 10^{-1}$ & 2.39$\times 10^{-1}$ & \\
\hline 
\end{tabular}
\end{center} 
\end{table*}

Relative  L$_3X$ double-hole to  L$_3$ single-hole configuration creation cross-sections are
compared in Table \ref{ratios} with those presented in reference~\cite{11}. Minor differences are due to shake-off values and to corrections for the atomic number, made in our calculations,  in the $P\left(  \text{L}_{i}\text{-L}_{j}X\right)$ values.

\begin{table*}
\begin{center}
\caption{Relative  Coster-Kronig and shake-off contributions for L$_3X$ double-hole to  L$_3$ single-hole creation cross-sections, $\sigma'_{L_3X}/\sigma'_{L_3}$, for gold (\%).}
\label{ratios}
\begin{tabular}{c|rrrr|r}\hline \hline
Spectator hole & 
\multicolumn{2}{c}{Coster-Kronig} &  
Shake-off & Total & Total \\
$X$ & L$_1$-L$_3X$ & L$_2$-L$_3X$ & L$_3X$ & This work & Oohashi~\cite{11} \\
\hline 
M$_{4}$ & 5.34 & 0.00 & 0.08 & 5.42 & 5.70 \\
M$_{5}$ & 7.07 & 0.00 & 0.10 & 7.17 & 7.60 \\
N$_{1}$ & 0.41 & 0.22 & 0.06 & 0.69 & 0.74 \\
N$_{2}$ & 0.17 & 1.50 & 0.09 & 1.76 & 1.80 \\
N$_{3}$ & 0.26 & 0.43 & 0.19 & 0.87 & 0.93 \\
N$_{4}$ & 0.64 & 1.60 & 0.39 & 2.63 & 2.70 \\
N$_{5}$ & 0.80 & 0.55 & 0.55 & 1.90 & 2.00 \\
N$_{6}$ & 0.51 & 0.07 & 1.20 & 1.77 & 1.80 \\
N$_{7}$ & 0.63 & 0.10 & 1.59 & 2.32 & 2.20 \\
O$_{1}$ & 0.08 & 0.04 & 0.26 & 0.39 & 0.39 \\
O$_{2}$ & 0.03 & 0.26 & 0.45 & 0.74 & 0.82 \\
O$_{3}$ & 0.05 & 0.07 & 1.03 & 1.15 & 1.20 \\
O$_{4}$ & 0.07 & 0.17 & 5.51 & 5.74 & 15.57 \\
O$_{5}$ & 0.08 & 0.06 & 9.69 & 9.83 & \\
P$_{1}$ & 0.01 & 0.00 & 10.07 & 10.08 & 6.50 \\
\hline
\end{tabular}
\end{center} 
\end{table*}

The X-ray production cross sections for the diagram lines can be written as 

\begin{equation}
\label{eq201}
\sigma^{\text{R}}\left( \text{L}_{3}\text{-N}_{k}\right) =\sigma_{\text{L}_{3}}^{\prime}  \dfrac{\Gamma \left( \text{L}_{3}\text{-N}_{k}\right) }{%
\Gamma _{\text{L}_{3}}^{\text{tot}}}
\end{equation}%
for the L$\beta _{2}$ ($k=5$) and L$\beta _{15}$ ($k=4$) lines, with $\sigma_{\text{L}_{3}}^{\prime}=432.45$ barn, and 
\begin{equation}
\label{eq202}
\sigma^{\text{R}}\left( \text{L}_{1}\text{-M}_{3}\right)  =\sigma_{\text{L}_{1}}^{\prime}   \dfrac{\Gamma \left( \text{L}_{1}\text{-M}_{3}\right) }{%
\Gamma _{\text{L}_{1}}^{\text{tot}}}\end{equation}%
for the L$\beta _{3}$ line, with $\sigma_{\text{L}_{1}}^{\prime}=139.55$ barn.
In equations (\ref{eq201}) and (\ref{eq202})  $\Gamma \left( \text{L}_{i}\text{-X}\right) $ is the
radiative width of the L$_{i}\text{-}$X line, and 
$\Gamma _{\text{L}_{i}}^{\text{tot}}$ is the
total width of the L$_{i}$ subshell. 

Assuming that satellite lines are caused by a single Coster-Kronig transition
and/or by shake-off processes, the X-ray production cross section for each line can be expressed as
\begin{eqnarray}
\sigma^{\text{R}}\left( \text{L}_{3}X\text{-}X\text{N}_{k}\right)  &=&\sigma_{\text{L}_{3}X}^{\prime}  
 \dfrac{\Gamma \left( \text{L}_{\text{3}}X\text{-}X\text{N}%
_{k}\right) }{\Gamma _{\text{L}_{\text{3}}X}^{\text{tot}}},  \label{eq004}
\end{eqnarray}
where  $\Gamma\left(  \text{L}_{\text{3}%
}X\text{-}X\text{N}_{\text{5}}\right)  $ is the radiative width of the
L$_{3}X$-$X$N$_{5}$ transition, and $\Gamma ^{\text{tot}}_{\text{L}_{3}X}$ is the
total width of the state with holes in the L$_3$ and $X$ subshells.

Widths for L$_1$ to N$_7$ levels were taken from Campbell and Papp~\cite{30}.
For the remaining one-hole levels, to our knowledge, no data are available in the literature.
 We used the following values $\Gamma_{\text{O}_1}=5.0$ eV, $\Gamma_{\text{O}_2}=3.8$ eV, $\Gamma_{\text{O}_3}=3.0$ eV, $\Gamma_{\text{O}_4}=2.4$ eV, $\Gamma_{\text{O}_5}=2.3$ eV, and $\Gamma_{\text{P}_1}=5.0$ eV. 
For the double-hole level widths we use ${\Gamma _{XY}^{\text{tot}}}={\Gamma _{X}^{\text{tot}}}+{\Gamma _{Y}^{\text{tot}}}$, where ${\Gamma _{X}^{\text{tot}}}$ and ${\Gamma _{Y}^{\text{tot}}}$ are single-hole level widths.

\begin{table*}
\begin{center}
\caption{Gold L$\beta_{15}$ and L$\beta_{2}$ satellite widths relative to L$\beta_{2}$ width (columns 2 and 3) and relative L$_3$ to L$_3X$ total level widths (column 4). In column 5, L$\beta_{15}$ and L$\beta_{2}$ satellite lines production cross sections relative to L$\beta_2$ diagram line are presented, Eq.(\ref{eq301}).}
\label{cross}
\begin{tabular}{clllll}\hline \hline
$X$ &  
$\frac{\Gamma (\text{L}_3X-X\text{N}_4)}{\Gamma (\text{L}_3-\text{N}_5)}$ & 
$\frac{\Gamma (\text{L}_3X-X\text{N}_5)}{\Gamma (\text{L}_3-\text{N}_5)}$ &
$\frac{\Gamma^\text{tot}(\text{L}_3)}{\Gamma^\text{tot}(\text{L}_3X)}$ &
$\frac{\sigma^\text{R} (\text{L}_3X-X\text{N}_4)}{\sigma^\text{R} (\text{L}_3-\text{N}_5)}$ & 
$\frac{\sigma^\text{R} (\text{L}_3X-X\text{N}_5)}{\sigma^\text{R} (\text{L}_3-\text{N}_5)}$ 
\\ 
\hline
M$_{4}$ & 0.118 & 1.024 & 0.72 & 0.0046 & 0.040 \\
M$_{5}$ & 0.120 & 1.021 & 0.72 & 0.0061 & 0.053 \\
N$_{1}$ & 0.144 & 0.983 & 0.39 & 0.0004 & 0.003 \\
N$_{2}$ & 0.112 & 0.998 & 0.46 & 0.0009 & 0.008 \\
N$_{3}$ & 0.160 & 0.952 & 0.52 & 0.0007 & 0.004 \\
N$_{4}$ & 0.125 & 0.701 & 0.57 & 0.0019 & 0.011 \\
N$_{5}$ & 0.132 & 0.830 & 0.59 & 0.0015 & 0.009 \\
N$_{6}$ & 0.114 & 0.759 & 0.94 & 0.0019 & 0.013 \\
N$_{7}$ & 0.099 & 0.890 & 0.94 & 0.0022 & 0.020 \\
O$_{1}$ & 0.112 & 1.008 & 0.5   & 0.0002 & 0.002 \\
O$_{2}$ & 0.111 & 0.873 & 0.6 & 0.0005 & 0.004 \\
O$_{3}$ & 0.092 & 0.998 & 0.7   & 0.0007 & 0.007 \\
O$_{4}$ & 0.105 & 1.109 & 0.7   & 0.004   & 0.044 \\
O$_{5}$ & 0.103 & 0.691 & 0.7   & 0.007 & 0.048 \\
P$_{1}$ & 0.111 & 1.000 & 0.5   & 0.006   & 0.053 \\ 
    \hline 
\end{tabular}
\end{center} 
\end{table*}
The L$\beta _{15}$ and L$\beta _{2}$ X-ray satellite line production cross sections ratio
relative to the L$\beta _{2}$ line are given by
\begin{equation}
\dfrac{\sigma ^{\text{R}}\left( \text{L}_{3}X-X\text{N}_{4,5}\right) }{\sigma
^{\text{R}}\left( \text{L}_{3}-\text{N}_{5}\right) }=\dfrac{\sigma _{\text{L%
}_{3}X}^{\prime }\dfrac{\Gamma \left( \text{L}_{3}X-X\text{N}_{4,5}\right) }{%
\Gamma _{\text{L}_{3}X}^{\text{tot}}}}{\sigma _{\text{L}_{3}}^{\prime }%
\dfrac{\Gamma \left( \text{L}_{3}-\text{N}_{5}\right) }{\Gamma _{\text{L}%
_{3}}^{\text{tot}}}}.\label{eq301}
\end{equation}
In equation (\ref{eq301})  $X=$ M$_{4}$,M$_{5}$ for the 
visible satellites and $X=$ N$_{1}$ to P$_{1}$ for the hidden
satellites. In their work, Oohashi \textit{el al}~\cite{11} assumed 
\begin{equation}
\dfrac{\dfrac{\Gamma \left( \text{L}_{3}X-X\text{N}_{5}\right) }{\Gamma _{%
\text{L}_{3}X}^{\text{tot}}}}{\dfrac{\Gamma \left( \text{L}_{3}-\text{N}%
_{5}\right) }{\Gamma _{\text{L}_{3}}^{\text{tot}}}}=1,
\end{equation}%
allowing for a further simplification of the equations. We
did not use here this simplification. Instead we explicitly calculated all the 
contributions for the radiative linewidths both in  diagram and satellite lines (Table~\ref{cross}).

\begin{figure*}
\centering
\includegraphics[width=16cm]{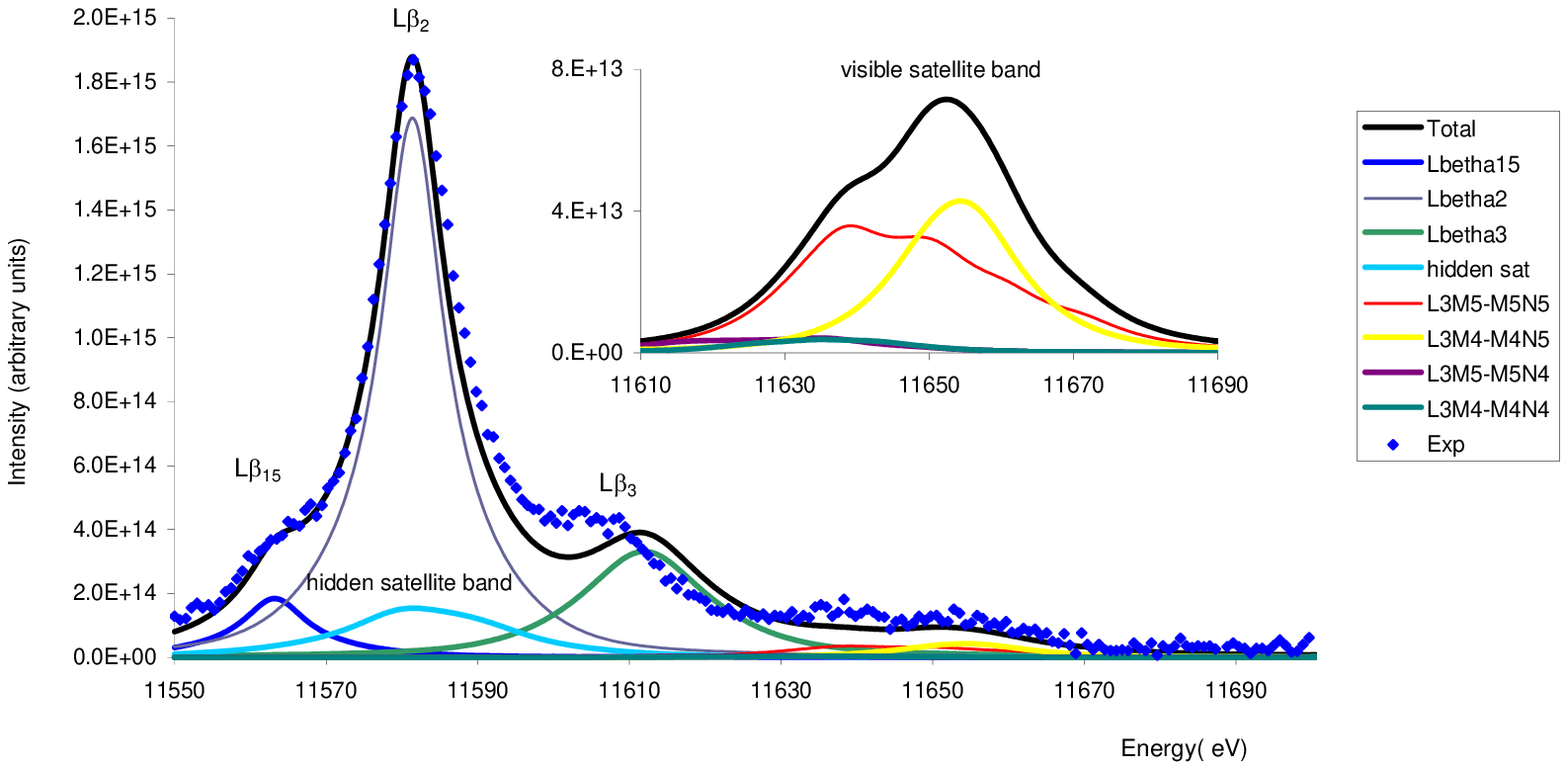}
\caption{Gold L X-ray  spectrum calculated in this work, including L$\beta_2$, L$\beta_3$, and L$\beta_{15}$
diagram lines and satellite bands, compared with the experimental data (diamonds)~\cite{11}. In the inset  L$\beta_2$, and L$\beta_{15}$  satellite lines are shown in a different intensity scale.}
\label{fig02}
\end{figure*}

The theoretical spectrum obtained using the
methods discussed above and assuming, for each line, a linear combination of a Gaussian and
a Lorentzian distribution,
is compared with the experimental data~\cite{11} in Fig.~\ref{fig02}. To allow for a better comparison,
 the experimental energy was shifted by -3.2 eV in order to superimpose the L$\beta_2$ line in both the theoretical
 and measured spectra. The discrepancies between the energy values, calculated in this work, of the diagram lines
  and the experimental ones are due to the neglect of correlation and other many-body corrections
 in our calculation \cite{26,27,33}.

\subsubsection{Calculation of gold L$\beta_{2}$ satellite band using Racah's
algebra}
\label{sec:3.3}

To assess the validity of the calculation of relative satellite intensities
using Racah's algebra in the case of tungsten, we used the same procedure for the
L$_{3}$M$_{4,5}$-M$_{4,5}$N$_5$ satellite bands
of gold, where a more precise calculation allows for a comparison.

For gold, angular momenta that result from unfilled outer shells are $j_{3}=\dfrac{1}{2}$ 
and $j_{3}^{\prime}=\dfrac{1}{2}$. Now, in what concerns equation (\ref{eq003}), for the L$_{3}$M$_{5}$-M$_{5}%
$N$_{5}$ transitions we have $J=\dfrac{1}{2},\ldots,\dfrac{9}{2}$ and
$J'=\dfrac{1}{2},\ldots,\dfrac{11}{2}$.

In order to compare the  spectra generated using the two different methods, we normalized
the total intensity of the L$_{3}$M$_{5}$-M$_{5}$N$_{5}$ satellite band to the same band
obtained with the method described in 3.2.2.

\begin{figure}
\centering
\includegraphics[width=8.5cm]{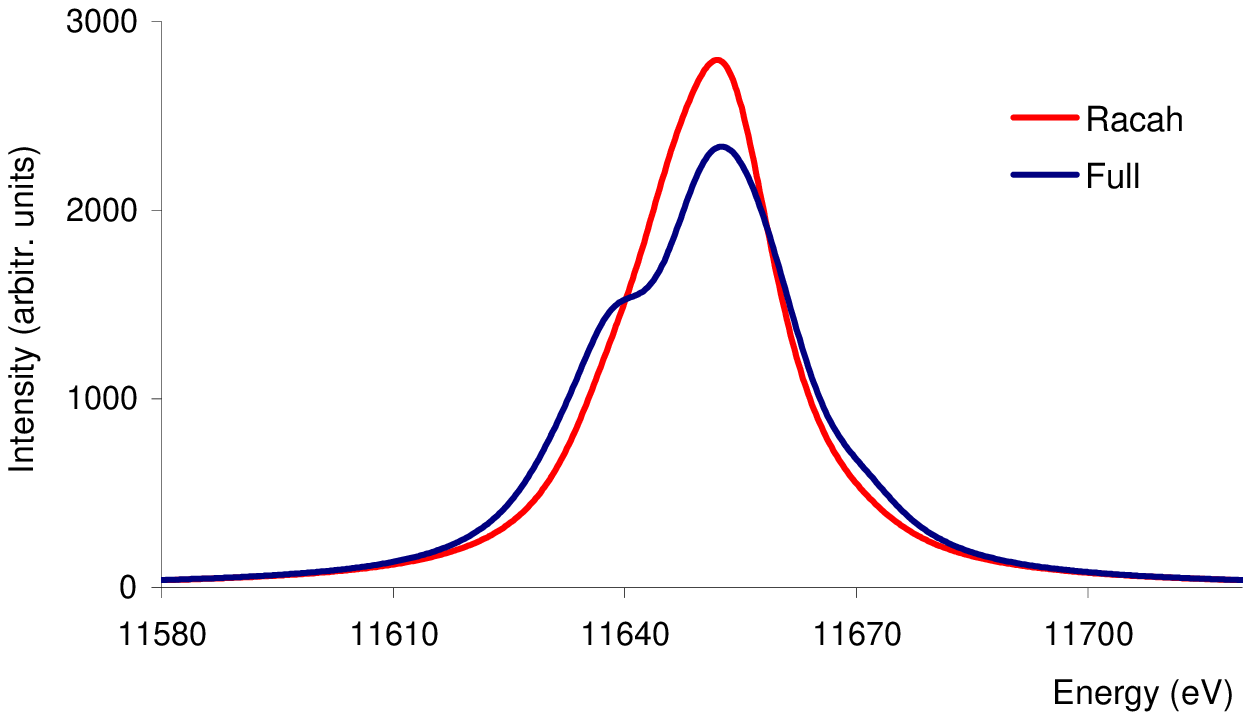}
\caption{Calculated  satellite spectrum of gold, including the
L$_{3}$M$_{4,5}$-M$_{4,5}$N$_{5}$ bands, using Multi-Configuration
Dirac-Fock energy values and intensities obtained with the Racah's algebra method - Racah,
and from the MCDF calculation - Full.} 
\label{fig03}
\end{figure}

The  L$_{3}$M$_{4,5}$-M$_{4,5}$N$_{5}$ satellite bands thus obtained are presented
in figure~\ref{fig03}. The agreement obtained allows us to conclude for the validity 
of the tungsten L$\beta_2$ spectrum we generated with relative intensities obtained with Racah's algebra.

\section{Discussion and conclusions}
\label{sec:4}

In this work we computed the theoretical L$\beta_{2}$ satellite band of
tungsten using MCDF relativistic transition energies and a statistical model
for the shape of the band.

Due to the large number of transitions involved, we were only
able to estimate the position and the shape of the the L$\beta_{2}$ satellite band, using a
method that employs Racah's algebra to find relative line intensities in this band.
Although the experimental values suffer from very poor statistics, our
results reproduce well the main features of the measured spectrum.

In the case of gold, we used the MCDF code to
compute both the transition energies and the transition probabilities of the
L$\beta_{2}$, L$\beta_{3}$, L$\beta_{15}$ diagram lines as well as the
L$\beta_{2}$  and L$\beta_{15}$ satellite lines, corresponding to spectator holes in the M$_{4}$
and M$_{5}$ subshells. These satellite lines are visible satellites. We also computed the
energies and transition probabilities for the hidden L$\beta_{2}$ and
L$\beta_{15}$ satellites. These are satellite lines, corresponding to
spectator holes in the N$_{i}$ ($i=1,7$), O$_{i}$ ($i=1,5$), and P$_{1}$
subshells, whose energies are such that they are superimposed on the
L$\beta_{2}$ diagram line.

\begin{table*}
\begin{center}
\caption{Energies and relative X-ray production cross sections of L$\beta_2$ plus L$\beta_{15}$ visible satellites for gold. Production cross sections are relative to L$\beta_2$ diagram line (a) and L$\beta_2$ diagram line plus hidden satellites (b) production cross sections, respectively.}
\label{rel_int}
\begin{tabular}{c|ccc|cccc}\hline \hline
& \multicolumn{3}{c|}{Energy (eV)} &   \multicolumn{4}{c}{Relative production cross section (\%)} \\
Spectator & 
This & \multicolumn{2}{c|}{Oohashi~\cite{11}} & 
\multicolumn{2}{c}{This work} & \multicolumn{2}{c}{Oohashi~\cite{11}} \\
hole & work & Theory & Experiment & (a) & (b) & Theory & Experiment \\
\hline 
       M$_{5}$ & 11646.9 & 11648.4 & 11641.7$\pm$ 1.1 & 5.9 & 4.6 & 5.5 & 4.1 $\pm$ 0.4 \\
       M$_{4}$ & 11654.0 & 11657.5 & 11658.7$\pm$ 1.1 & 4.4 & 3.5 & 4.2 & 4.1 $\pm$ 0.8 \\
\hline
\end{tabular}
\end{center} 
\end{table*}

The values found in this work for the ratio of L$_{3}$M$_{4,5}$-M$_{4,5}$N$_{4,5}$ to L$_{3}$-N$_{5}$ 
X-ray production cross sections are presented in Table~\ref{rel_int}.
For comparison with the experiment, the ratios  of the
L$_{3}$M$_{4,5}$-M$_{4,5}$N$_{4,5}$ to L$_{3}$-N$_{5}$ plus hidden satellites production cross sections were also computed.
These ratios are compared with Oohashi \textit{et al}~\cite{11} theoretical and experimental values on the same table. We note that in the latter calculation, L$_{3}$M$_{4,5}$-M$_{4,5}$N$_{4}$ satellite line production cross sections were
not taken in account. Also, transition yields for satellite and diagram lines are taken by this authors as equal.

We estimate that the uncertainty in our calculation of the relative X-ray production cross sections
is of the order of 20\%, due
mainly to the uncertainty in the L$_{1}$- and L$_{3}$-subshells ionization cross sections which are 20\%
and 10\%, respectively.

The theoretical spectrum for gold obtained in this work agrees very
well with experiment. Taking in account the uncertainties, the satellite bands relative intensities found in this
work are consistent with the measured values of Oohashi \textit{et al}~\cite{11}.

\section*{Acknowledgments}
\label{sec:5}

We thank Doctors H. Oohashi, A. M. Vlaicu, and Y.
Ito for their kind collaboration in making available to us
their L$\beta$ sprectra for gold.

This research was partially supported by the FCT pro\-jects
POCTI/FAT/50356/2002 and \ POCTI/0303/2003 \ (Portugal),
financed by the European Community Fund FEDER.
Laboratoire Kastler Brossel is Unit{\'e} Mixte de Recherche du CNRS
n$^{\circ}$ C8552

%

%


\end{document}